\documentclass{aastex}

\shorttitle{Aharonian et al.}
\shortauthors{The Crab~Nebula Energy Spectrum }

\begin{document}

\title{The Energy Spectrum of TeV Gamma-Rays from the Crab~Nebula 
as measured by the HEGRA system of imaging air \v{C}erenkov telescopes}

\author{
F.A.~ Aharonian\altaffilmark{1}, 
A.G.~Akhperjanian\altaffilmark{7},
J.A.~Barrio\altaffilmark{2,3},
K.~Bernl\"ohr\altaffilmark{1,4},
H.~Bojahr\altaffilmark{6},
I.~Calle\altaffilmark{3},
J.L.~Contreras\altaffilmark{3},
J.~Cortina\altaffilmark{3},
S.~Denninghoff\altaffilmark{2},
V.~Fonseca\altaffilmark{3},
J.C.~Gonzalez\altaffilmark{3},
N.~G\"otting\altaffilmark{4},
G.~Heinzelmann\altaffilmark{4},
M.~Hemberger\altaffilmark{1},
G.~Hermann\altaffilmark{1},
A.~Heusler\altaffilmark{1},
W.~Hofmann\altaffilmark{1},
D.~Horns\altaffilmark{4},
A.~Ibarra\altaffilmark{3},
R.~Kankanyan\altaffilmark{1,7},
M.~Kestel\altaffilmark{2},
J.~Kettler\altaffilmark{1},
C.~K\"ohler\altaffilmark{1},
A.~Kohnle\altaffilmark{1},
A.~Konopelko\altaffilmark{1},
H.~Kornmeyer\altaffilmark{2},
D.~Kranich\altaffilmark{2},
H.~Krawczynski\altaffilmark{1},
H.~Lampeitl\altaffilmark{1},
A.~Lindner\altaffilmark{4},
E.~Lorenz\altaffilmark{2},
F.~Lucarelli\altaffilmark{3},
N.~Magnussen\altaffilmark{6},
O.~Mang\altaffilmark{5},
H.~Meyer\altaffilmark{6},
R.~Mirzoyan\altaffilmark{2},
A.~Moralejo\altaffilmark{3},
L.~Padilla\altaffilmark{3},
M.~Panter\altaffilmark{1},
R.~Plaga\altaffilmark{2},
A.~Plyasheshnikov\altaffilmark{1},
J.~Prahl\altaffilmark{4},
G.~P\"uhlhofer\altaffilmark{1},
G.~Rauterberg\altaffilmark{5},
A.~R\"ohring\altaffilmark{4},
V.~Sahakian\altaffilmark{7},
M.~Samorski\altaffilmark{5},
M.~Schilling\altaffilmark{5},
D.~Schmele\altaffilmark{4},
F.~Schr\"oder\altaffilmark{6},
W.~Stamm\altaffilmark{5},
M.~Tluczykont\altaffilmark{4},
H.J.~V\"olk\altaffilmark{1},
B.~Wiebel-Sooth\altaffilmark{6},
C.~Wiedner\altaffilmark{1},
M.~Willmer\altaffilmark{5},
W.~Wittek\altaffilmark{2}}

\altaffiltext{1}{Max Planck Institut f\"ur Kernphysik,
Postfach 103980, D-69029 Heidelberg, Germany}
\altaffiltext{2}{Max Planck Institut f\"ur Physik, F\"ohringer Ring
6, D-80805 M\"unchen, Germany}
\altaffiltext{3}{Universidad Complutense, Facultad de Ciencias
F\'{\i}sicas, Ciudad Universitaria, E-28040 Madrid, Spain }
\altaffiltext{4}{Universit\"at Hamburg, II. Institut f\"ur
Experimentalphysik, Luruper Chaussee 149,
D-22761 Hamburg, Germany}
\altaffiltext{5}{Universit\"at Kiel, Institut f\"ur Experimentelle und 
Angewandte Physik,
Leibnizstra{\ss}e 15-19, D-24118 Kiel, Germany}
\altaffiltext{6}{Universit\"at Wuppertal, Fachbereich Physik,
Gau{\ss}str.20, D-42097 Wuppertal, Germany}
\altaffiltext{7}{Yerevan Physics Institute, Alikhanian Br. 2, 375036
Yerevan, 
Armenia}

\begin{abstract}
The Crab Nebula has been observed by the HEGRA ({\it High-Energy Gamma-Ray Astronomy}) 
stereoscopic system of imaging air \v{C}erenkov telescopes (IACTs) for a total 
of about 200~hrs during two observational campaigns: from September~1997 to March~1998 
and from August~1998 to April~1999. The recent detailed studies of system performance 
give an energy threshold and an energy resolution for $\gamma$-rays of 500~GeV and 
$\sim 18$\%, respectively. The Crab energy spectrum was measured with the HEGRA IACT 
system in a very broad energy range up to 20~TeV, using observations at zenith angles 
up to 65 degrees. The Crab data can be fitted in the energy range from 1 to 20~TeV by 
a simple power-law, which yields 
$\rm dJ_{\gamma}/dE = (2.79\pm 0.02 \pm 0.5) \cdot 10^{-7} 
(\rm \frac{E}{\rm 1\, TeV})^{-2.59\pm 0.03 \pm 0.05} 
ph\,m^{-2}\,s^{-1}\,TeV^{-1}$. 
The Crab~Nebula energy spectrum, as measured with the HEGRA IACT system, agrees within 
15\% in the absolute scale and within 0.1 units in the power law index with the latest 
measurements by the Whipple, CANGAROO and CAT groups, consistent within the statistical 
and systematic errors quoted by the experiments. The pure power-law spectrum of TeV 
$\gamma$-rays from the Crab~Nebula constrains the physics parameters of the nebula 
environment as well as the models of photon emission.
\end{abstract}


\section{Introduction}
\label{intro.sec}

The Crab~Nebula has been observed and studied over an enormously broad photon energy range 
embracing the radio, optical, and X-ray bands, as well as, high energy $\gamma$-ray region 
up to hundreds of TeV. The various theoretical scenarios of photon emission are primarily 
based on the Synchro Compton model (Gould 1965) which combines the synchrotron and inverse 
Compton (IC) emissions from high-energy electrons, which are accelerated up to $\sim 100$~TeV 
and interact with the magnetic field and the low frequency seed photons within the Nebula 
(de Jager \& Harding 1992; Atoyan \& Aharonian 1996; de Jager et al. 1996; Hillas et al. 1998). 
The predicted IC spectrum in the TeV energy domain appears to be very sensitive to the model 
parameters, such as the value of the magnetic field, the nature of the seed photons, the maximum 
energy of electrons, etc. Although IC scenarios of the photon emission are widely believed to 
be appropriate for the Crab~Nebula one can not exclude the possible contribution of $\gamma$-ray 
fluxes from $\pi^\circ$-decay (see Atoyan \& Aharonian 1996, Bednarek \& Protheroe 1997). Thus, 
a measurement of TeV Crab~Nebula spectrum sets major constraints on theoretical expectations 
and precise spectral measurements allow to fix the model parameters. 

The imaging air \v{C}erenkov technique was successfully used for observations of the Crab~Nebula 
in TeV $\gamma$-rays. Since the time of detection at a 9$\sigma$ confidence level by the Whipple 
group (Weekes et al. 1989) a number of observations of the Crab~Nebula have been made at TeV 
energies (for reviews see Ong 1998; Catanese \& Weekes 1999). By now the Crab Nebula is 
established as the standard candle of steady TeV $\gamma$-ray emission. The optical Nebula of 
the Crab has an angular extension of about 6~arc~min (Hester~et~al.~1995). The standard analysis 
for the HEGRA system of IACTs provides an angular resolution of about $0.1^\circ$ (see Figure~1). 
In the present analysis the Crab~Nebula was assumed to be as a $\gamma$-ray point source (see 
Figure~2). The detailed mapping of the TeV $\gamma$-rays from the Crab~Nebula undertaken by 
Hofmann~(1999) provided an upper limit for the Crab angular extension in TeV $\gamma$-rays of 
$\sim1.5$~arcmin. No pulsed emission has been seen so far from the Crab (Gillanders et al. 1997; 
Burdett et al. 1999; Aharonian~et~al. 1999a). Vacanti~et~al (1991) provided a first measurement of 
the energy spectrum of TeV $\gamma$-ray emission from the Crab~Nebula using 10~m Whipple telescope 
data and derived a power-law spectrum index of 2.4 and a differential $\gamma$-ray flux at 
400~GeV of $\rm 2.5\cdot 10^{-10}\, photons \, cm^{-2} \, s^{-1}$. Note that the absolute calibration 
of individual imaging air \v{C}erenkov telescopes is quite difficult because there is no ``test'' beam 
of TeV $\gamma$-rays. The uncertainties in absolute calibration propagate to uncertainties of 
estimated $\gamma$-ray fluxes. Different methods of telescope calibration have been developed 
recently in order to reduce systematic errors which could influence the estimate of the absolute 
$\gamma$-ray flux and the slope of energy spectrum (see e.g., Frass et al. 1997). The energy 
spectrum measurements heavily rely on Monte Carlo simulations of the telescope response. Thus in 
the past the analysis of the same observational data using various methods of telescope calibration 
revealed, not infrequently, very different estimates of the telescope energy threshold and of the TeV 
$\gamma$-ray fluxes. These estimates may vary by more than a factor of 2.  Thus, based on data taken 
with the prototype HEGRA imaging \v{C}erenkov telescope (CT2), Konopelko~et~al. (1996) detected the 
TeV $\gamma$-ray signal from the Crab~Nebula with a signal-to-noise ratio of 10$\sigma$ and derived 
from the data a power-law index of the energy spectrum of $2.7\pm 0.1$ and an integral flux of TeV 
$\gamma$-rays above 1~TeV of $\rm 8(\pm 1)_{\rm Stat}(\pm 2.4)_{\rm Syst} \cdot 10^{-12}\, photons\, 
cm^{-2} \, s^{-1}$. However, later on, the detailed treatment of the telescope hardware (optical 
smearing, photon-to-photoelectrons conversion efficiency, etc) allowed a more precise 
estimate of the TeV $\gamma$-ray flux using the single telescope data, as $\rm 1.5 (\pm0.2)_{Stat}
(+1.0-0.5)_{Syst} \cdot 10^{-11}\, photons \, cm^{-2} \, s^{-1}$ above 1~TeV (Petry et al. 1996). 

Significant improvements in the telescope hardware as well as in the simulations and the data analysis 
(Fegan, 1997) recently provided measurements of the Crab~Nebula energy spectrum over the energy range 
from 200~GeV up to 50~TeV by several groups, Whipple (Hillas~et~al. 1998), CANGAROO (Tanimori~et~al. 
1998), and CAT (see Barrau 1998) exploiting the imaging \v{C}erenkov technique. An additional 
measurement was provided by a low energy air shower array (Amenomori et al. 1999). However the 
uncertainties of the $\gamma$-ray flux estimates as well as of the energy spectrum slope remain rather 
large, in particular at the high energy end of the measurements i.e. beyond 10~TeV. For example the 
Tibet data (Amenomori~et~al.~1999) show significantly higher (by a factor of 2) $\gamma$-ray fluxes 
compared with the results obtained using the IACTs in the energy range from 3 to $\sim 18$~TeV, and are 
in favor of a gradual steepening of the spectral slope at high energies. Thus additional precise 
measurements of the Crab Nebula energy spectrum in TeV $\gamma$-rays are of great importance. The HEGRA 
system of IACT provides such data. It was primarily designed for detailed spectral measurements in the 
TeV energy domain utilizing the advantages of {\it stereoscopic} observations. Stereo imaging gives 
several advantages for spectral studies, compared to a single telescope: {\it (i)} direct measurement of 
the shower impact parameter with an accuracy better than 10~m {\it (ii)} good energy resolution of 18\% 
{\it (iii)} wide dynamic range from 500~GeV to 20~TeV {\it (iv)} extended abilities for systematic studies 
using several images for an individual shower. The detailed systematic studies for the spectrum evaluation 
technique have been recently made using Mrk~501 1997 observational data (Aharonian~et~al. 1999b, 1999c). The 
performance of the system was discussed by Konopelko~et~al. (1999a). Here we present Crab Nebula data 
taken with the HEGRA telescope system in 1997/1998 and 1998/1999 observational campaigns. The data were 
analyzed using a new technique of energy spectrum evaluation for the {\it stereoscopic observations}. 
We also discuss the physics implications of the present results for the modeling of the TeV $\gamma$-ray 
emission from the Crab~Nebula.   

\section{Observational data}
\label{data.sec}

The Crab Nebula was extensively observed with the HEGRA IACT system in two observing seasons from 
September~1997 to March~1998 and from October~1998 to April~1999. The observations were made with the 
stereoscopic system of IACTs which are located on La Palma, Canary Islands (Aharonian~et~al. 1999c). 
Each of the telescopes consists of a 8.5~$\rm m^2$ reflector focussing the \v{C}erenkov light onto a 
photomultiplier tube camera. The 271 photomultipliers in the camera were arranged in a hexagonal matrix 
covering a field of view with a radius of $2.15^\circ$. The telescope camera was triggered when the 
signal in two next neighbors of the 271 photo multiplier tubes exceeded a threshold of 8 photoelectrons, 
and the system readout started when at least two telescopes were triggered by \v{C}erenkov light from 
an air shower. The detection rate was 12.6~Hz near the Zenith in December~1997 and dropped down to about 
10~Hz in December~1999 for the 4-telescope system due to aging of the PMTs and reduced mirror reflectivity. 

The Crab Nebula was observed in a ``wobble mode''; i.e., the telescopes were pointed in Declination 
$\pm 0.5^\circ$ away from the nominal Crab Nebula position (the sign of the angular shift was altered from 
one run of 20 min to the next). This is useful for continuous monitoring of the cosmic-ray background 
because it positions the OFF-source region symmetric to the camera center, and $1^\circ$ apart from the 
ON-source region. Observations of the Crab Nebula at zenith angles up to 50 degree were made with 4 
telescopes from 1997 September 1 to 1998 March 29, for a total of 82.5 hr of data taken at good weather. 
Through the fire at the HEGRA site one of the telescopes was damaged and was out of operation for a month 
in October-November 1997. At that time, Crab Nebula observations were made with only three telescopes in 
the system, providing an event rate of 10.3~Hz near the zenith. Due to unstable weather and a substantial 
amount of dust which came from Sahara desert to the island, the average detection rate in 1998 
February-March was reduced down to 10.7~Hz in observations near the zenith. At the beginning of 1998/1999 
observational period 4 telescopes were operational. Since October~1998 the HEGRA collaboration operates 5 
telescopes. However, for technical reasons one telescope (CT2) was out of operation since December~1998 
until the end of the first observation campaign (April~1999). During the last observational period (since 
August~1999) observations at zenith angles less than $50^\circ$ were taken for about 76.1~hrs. In addition, 
the observations at large zenith angles (LZA) ($50^\circ < \theta < 65^\circ$), were carried out for a total 
of 24~hrs in order to study the performance of the telescope system at LZA and to extend the measurements 
of the Crab Nebula energy spectrum beyond 10~TeV. The total exposure times for the three periods are 
summarized in Table 1.

Only data taken under good weather conditions were used in the analysis. In order to exclude data taken 
under less than optimal telescope performance conditions the entire database has been checked very carefully 
as follows: First, each night the compressed protocols of the system performance were transfered to one of 
the collaboration host institutes where they were scanned by software tools which closely monitor the status 
of the telescopes' hardware (single pixel rates, trigger rates of system telescopes, tracking accuracy, 
etc). This information was accumulated in a corresponding database which was used afterwards for a standard 
data reduction procedure. The final condensed data file for each particular run contains all the information 
needed for data analysis. In addition a specific software tool was developed which allows to control 
{\it a posteriori} for each data run {\it (i)} the system trigger rate, taking into account the zenith angle 
dependence {\it (ii)} the angular shape of the cosmic ray images, tested by a $\chi^2$--criterion for the 
deviation of the {\it mean scaled Width} distribution for a single run from the corresponding average 
distribution filled over an extended sample of runs {\it (iii)} the flatness of the $\theta^2$--distribution 
for the isotropic cosmic ray images over the full field of view {\it (iv)} the image {\it Size} distributions 
for each individual telescope. 

\section{Analysis}
\label{ana.sec}

The {\it stereoscopic imaging} analysis of the data is based on the geometrical reconstruction of the 
shower arrival direction and the shower core position in the observation plane, as well as on the joint 
parameterization of the shape of the \v{C}erenkov light images. The simultaneous registration of several
($\geq$2) \v{C}erenkov light images from an air shower provides an angular resolution of $\sim 0.^\circ 1$ 
for $\gamma$-ray showers. For each individual shower, stereoscopic observations allow to determine the 
position of the shower axis. Thus, at first only air showers within a certain impact distance $\rm R_0$ from 
the center of the telescope system were selected. The limiting upper radius of $\rm R_0 = 200\,m$ was used 
for zenith angles less than 50 degrees, and a significantly larger radius of 400 m for the large zenith angle 
observations ($>50$~degrees). The effective collection area in observations at LZA dramatically increases at 
high energies, far beyond the limiting radius of 200~m (Konopelko et al. 1999b). For the data taken at zenith 
angles up to $50^\circ$ an orientation cut $\theta^2<0.05\, \rm [deg^2]$ was applied, where $\theta^2$ is the 
squared angular distance of the reconstructed source position from the true source position. In addition the 
data were analyzed using the {\it mean scaled Width} parameter, $<\tilde{w}>$. To compensate for the dependence 
of the image shape on primary shower energy and distance from shower core to the telescope (impact parameter), 
the standard parameter {\it Width} (Fegan, 1997) ($w^k$), calculated for each telescope, is scaled according 
to the Monte Carlo predicted values, for $\gamma$-rays $<w>^k_{ij}$, taken for the corresponding bin of 
reconstructed distance from the telescope to the shower core (i) and for the corresponding bin of image size 
(total number of photoelectrons in the image)(j) (Aharonian~et~al. 1999b, 1999c). The {\it mean scaled Width} 
parameter is defined for each individual shower as follows
\begin{equation}
< \tilde w > = 1/N \sum_{k=1}^{N} w^k/<w>^k_{ij}
\end{equation} 
where $N$ is the number of triggered telescopes. This parameter was introduced in order to provide an almost 
constant $\gamma$-ray acceptance over the dynamic energy range of the telescope system. The optimum cut on 
{\it mean scaled Width} is about 1.1, which gives a $\gamma$-ray acceptance of $\sim 60$\% at Small Zenith 
Angle (SZA). However, for a precise determination of $\gamma$-ray spectra, a loose cut on {\it mean scaled Width} 
($<\tilde w> < 1.2$) has been so far used in the data analysis in order to maximize the $\gamma$-ray acceptance 
and to minimize systematic errors related to cut efficiencies. Thus, the second $\gamma$-ray selection criterion 
was $<\tilde{w}>\, \leq 1.2$. This set of cuts was found to be optimal for spectrum studies (Aharonian~et~al. 
1999b, 1999c). These {\it loose} analysis cuts provide a Crab Nebula $\gamma$-ray rate of 83~$\gamma$s/hr at SZA 
(less that $25^\circ$) for the 5-IACT system. The corresponding energy threshold of the $\gamma$-rays is about 
500~GeV. For the LZA data the looser orientation cut of $\theta^2 < 0.1\, \rm [deg^2]$ was used because of the 
lower accuracy of the arrival direction reconstruction for the $\gamma$-ray showers. In order to improve the 
cosmic ray rejection in observations at LZA an additional parameter, {\it mean scaled Length}, $<\tilde l>$ was 
used, which is defined by analogy with $<\tilde w>$ (Konopelko~et~al. 1999b). These two parameters, $<\tilde w>$ 
and $<\tilde l>$, can be used for calculating a Mahalanobis distance, MD (Mahalanobis 1963), in two-dimensional 
space as
\begin{equation}
\rm MD = ((1-<\tilde w>)^2/\sigma^2_{<\tilde w>}+(1-<\tilde l>)^2/\sigma^2_{<\tilde l>})^{1/2}
\end{equation}
where $\sigma_{<\tilde w>}$ and $\sigma_{<\tilde l>}$ are the standard deviations for the corresponding
distributions of $<\tilde w>$ and $<\tilde l>$. The optimum value of the MD cut for LZA is found to be 1.5. 
Note that this analysis improves the enhancement factor by $\simeq 30$\% (it gives $\sim 50$\% acceptance of 
$\gamma$-rays) in observations at LZA, whereas it gives only marginal improvement for the data 
taken at SZA. The Crab Nebula $\gamma$-ray rate in observations at LZA ($60^\circ$) is about 16~$\gamma$'s/hr 
with a corresponding energy threshold of $\sim$5~TeV. Note that SZA observations give a $\gamma$-ray rate at 
high energies (above 3~TeV) of $\sim 8\,\gamma$s/hr. A summary of the data is shown in Table~2.

The observations of the Crab Nebula have been made during 6 periods which differ in the system configuration, 
reflectivity of the mirrors, light reflection by the pixel funnels and camera protecting plate {\it etc}. All 
that affects the hardware event rate of the telescope system, $\rm R_{exp}$. These changes of system performance 
were implemented in the Monte Carlo simulations. Assuming the standard chemical composition of the primary cosmic 
rays (Wiebel, 1994) the calculated detection rates, $\rm R_{MC}$, were adjusted to the measured rates 
(see Table~3).  

The collection areas, as a function of energy and zenith angle, for $\gamma$-ray showers have been inferred 
from Monte Carlo simulations (Konopelko et al. 1999a). The rms error of the energy determination is 
$\rm \Delta E/E \sim0.18$. The Monte Carlo studies show that for a good energy resolution of 18\% this 
approach does not distort the initial spectrum shape. The collection area for $\gamma$-rays rises very 
quickly in the energy range near the energy threshold of the telescope system, which is 500 GeV, whereas it 
is almost constant at the energies above $\geq 3$ TeV. Even slight variations of the trigger threshold could 
lead to noticeable systematic changes in the predicted spectral behavior in the energy range of $\sim 0.5-1$ TeV. 
This effect leads to a noticeable probability for ``sub-threshold'' triggers. In addition, the trigger level 
for different camera pixels is slightly different even after very accurate adjustment of the high voltage 
using the calibration laser runs. Measurements of the trigger setting for a number of camera pixels revealed 
variation in the trigger threshold of order $10$\%. These variations were implemented into the simulations 
in order to estimate the corresponding systematic error of the energy spectrum at energies below 1~TeV.
The fine tuning of the Monte Carlo simulations with respect to the IACT system data provided measurements of 
the flux of the cosmic ray protons in the energy range from 1.3 to 10~TeV (Aharonian~et~al. 1999d). The proton 
fluxes as measured by the HEGRA IACT system are perfectly consistent with the results of a bulk of satellite 
experiments held in this energy range. 

The procedure for the evaluation of the energy spectrum using the {\it stereoscopic} observations was discussed 
in detail by Aharonian~et~al. (1997); Hofmann (1997), and, more recently in Aharonian~et~al. (1999c). In the 
stereoscopic observations the impact distance of the shower axis to a system telescope can be measured with an 
accuracy $\leq 10$ m. The energy E of a $\gamma$-ray shower is defined by interpolation over the ``size'' 
parameter $\rm S$ (total number of photoelectrons in \v{C}erenkov light image) at a fixed impact distance $R$, 
as $\rm E = f_{MC}( S,R,\theta)$, where $\theta$ is the zenith angle and $\rm f_{MC}$ is a function obtained 
from Monte Carlo simulations. Note that the Monte Carlo simulations used here include the sampling of detector 
response in great detail (Hemberger 1998). The energy distribution for the ON- and OFF-source events, after the 
orientation and shape image cuts, were histogrammed over the energy range from 500 GeV to 30 TeV with 8 bins per 
decade. The $\gamma$-ray energy spectrum was obtained by subtracting ON- and OFF-histograms and dividing the 
resulting energy distribution by the corresponding collection area and the $\gamma$-ray acceptance. In the 
present Crab Nebula analysis the energy spectrum measurements were extended up to large zenith angles ($65^\circ$). 
The data were processed independently for each of the four zenith angle bins: $\rm (0^\circ - 25^\circ), 
(25^\circ - 40^\circ), (40^\circ - 50^\circ), (50^\circ - 65^\circ)$. The corresponding effective collection 
areas as well as the cut efficiencies were calculated as a function of the zenith angle. First, the energy 
spectra were derived for all zenith angle bins independently. Note that the spectra evaluated at different 
zenith angles are in a good agreement. For the final energy spectrum the different zenith angle bins were joined 
according to the prescription: 
\begin{eqnarray}
dJ_\gamma^i/dE = \sum_{j=1}^4 w_j (dJ_\gamma^i/dE)_j \Theta (E^i-E^j_{th}), \nonumber \\ 
w_j = t_j/t_0, \, i=1,n; 
\end{eqnarray}
where $dJ^i_\gamma/dE$, $(dJ^i_\gamma)_j/dE$ are the differential energy spectra at energy $E^i$ as measured over 
all zenith angle ranges, and for the particular zenith angle bin (j), respectively. $E_{th}^j$ is an estimated energy 
threshold for the zenith angle bin $j$, $t_j$ is the observation time for the $j$-bin on the zenith angle, and 
$t_0$ is the total observation time. 
The first three zenith angle bins were joined using the time dependent weights 
$\rm w_j = t_j/t_0$. Finally the spectrum measured in the zenith angle range of $\rm (0^\circ - 50^\circ)$ was 
combined with the spectrum derived from large zenith angle data, $\rm (50^\circ - 65^\circ)$ using the weights based 
on the estimated statistical errors for both spectra. Such procedure takes into account the advantageous $\gamma$-ray 
rate in LZA observations at high energies.  

The statistics of the $\gamma$-rays from the Crab Nebula provides a measurement of the energy spectrum up to 
a few tens of TeV. However, detection of \v{C}erenkov light images with extremely large amplitudes - several 
thousands of ph.e. - is complicated by the nonlinearity in the PMT response as well as by the saturation in 
the 8 bit Flash-ADC readout. Measurements of the photomultiplier response under high light loads 
over the extended sample of the EMI 9073 PMTs gave a calibration function which was used to correct 
the image amplitudes. The readout of the HEGRA IACT is based on the sampling of \v{C}erenkov light time 
impulse by the 16 FADC bins of $\sim 8$ ns each (Hess et al. 1998). The time pulses from the air showers 
with a full width at half maximum of a few ns were widened using an electronic scheme in order to fit into 
several FADC bins for the accurate measurement of the time profile. The smoothing of the FADC signal 
was unfolded back to the impulse, which almost always fits 2 FADC bins. The calibrated amplitude, summed 
over two FADC bins, is used as a measure of the pixel signal. For the high energy air showers the FADC 
signals run into saturation and the simple unfolding procedure fails. For such pulses the initial amplitude 
is reconstructed using the additional calibration function obtained by simultaneous measurements of light 
flashes with FADCs and a 14 bit ADC. This procedure drastically extends the dynamic range of the FADC readout.

To avoid the saturation problem one might only use images detected from air showers at large impact distances 
from the telescope system (e.g. beyond 150 m). The size of these images is very small even for high energy 
events because of the low \v{C}erenkov light density far off the shower axis. However these images are very 
often truncated by the camera edge and do not allow a proper reconstruction of the shower impact point and 
of the shower energy. This effect becomes less important in observations at LZA because of the high shower 
maximum height (the images shrink to the camera center). In the present analysis the maximum impact distance 
of the shower core from the center of the system was extended up to 400~m for observations at LZA. Observations 
at LZAs permit measurements of the energy spectrum far beyond 10~TeV. The images of $\gamma$-ray air showers 
observed at LZAs have small {\it Size} and are not influenced by the saturation effect. 

\section{Results}
\label{res.sec}

We have observed the Crab Nebula extensively in two observational seasons with the HEGRA IACT system. 
The HEGRA system of 5 IACTs currently has a sensitivity which allows the detection of a $\gg$5$\sigma$  
signal from the Crab Nebula within 1~hr of observation time (see Figure~1). The integral $\gamma$-ray fluxes 
measured during the different observational periods are consistent within the estimated statistical and 
systematic error (see Table~3). The differential energy spectrum of the Crab Nebula has been derived from the 
HEGRA data for two observational campaigns using recently developed advanced techniques for the measurements of 
the spectrum using {\it stereoscopic data} taken at small and large zenith angles. The Crab Nebula differential 
energy spectrum derived from SZA data matches quite well the spectrum derived at LZA (see Figure~3). The 
$\gamma$-ray rate measured at energies above 10 TeV in observations at LZAs exceeds the corresponding rate 
measured at SZA by a factor of 3. The LZA data are not affected by saturation effects. At 3.7$\sigma$ confidence 
level 27 $\gamma$-ray events from the Crab Nebula were detected in the highest energy bin from 17.8 to 23.7~TeV 
(see Figure~4). One may expect that a number of $\gamma$-ray events in the highest energy bin are spilled over 
from the lower energies. However, given the good energy resolution of 20\% and the power law energy spectrum, 
such effect is very small and is compensated almost by the backwards influx of the $\gamma$-rays from the energies 
above (see e.g., Aharonian~et~al. 1995). Finally, the simulations show that spilling over of low energy $\gamma$-rays 
does not influence the resulting fluxes measured at the upper end of the power law spectrum.     
       
The analysis for the different system configurations as well as for different trigger threshold 
values gives a differential energy spectrum of the Crab Nebula measured at zenith angles up to $60^\circ$
\begin{equation}
\rm dJ_{\gamma}/dE = (2.79\pm 0.02 \pm 0.5) \cdot 10^{-7} (\frac{E}{1\, TeV})^{-2.59\pm 0.03 \pm 0.05} 
ph\,m^{-2}\,s^{-1}\,TeV^{-1}
\end{equation}
The statistical and systematic errors are also given. The final Crab Nebula spectrum as measured by the 
HEGRA collaborations is shown in Figure~4. The measured $\gamma$-ray fluxes are given in Table~4.
The Crab Nebula energy spectrum is best fitted by a pure power law in the energy range 1-20~TeV. It does not 
exclude a possible slight steepening of the energy spectral usually predicted 
by inverse Compton modeling of TeV $\gamma$-ray emission. A fit with a logarithmic steepening of the 
power law spectrum gives the following result
\begin{equation}
\rm dJ_{\gamma}/dE = (2.67 \pm 0.01 \pm 0.5) \cdot 10^{-7} (\frac{E}{1\, TeV})^{-2.47 \pm 0.1 
\pm 0.05 - (0.11 \pm 0.10)\, log(E)} ph\,m^{-2}\,s^{-1}\,TeV^{-1}
\end{equation}
Such a fit indicates the slight flattening of the spectrum at low energies as predicted by the IC calculations.
However, the change of the energy spectrum slope is within the current statistical and systematic errors, and 
the data for the overall energy range are consistent with a simple power law fit in 0.5-20 TeV. The HEGRA Crab 
Nebula data match well the recent 20~TeV data published by the CANGAROO group (Tanimori et al. 1998), and are 
consistent with a flat power law index of $\sim 2.5$ beyond 20 TeV. The compilation of the world data is given 
in Figure~5. All data are consistent within statistical and systematic errors, except possibly for the Tibet 
data which show relatively higher fluxes. 

\section{Astrophysics implications}

The TeV energy spectrum of the Crab Nebula as measured by the HEGRA system of IACTs is consistent with the 
expectations for the TeV $\gamma$-ray emission from pulsar-driven Nebulae (plerions). According to this 
scenario the ultra relativistic electrons, accelerated in the pulsar wind shock, produce TeV $\gamma$-rays 
through the IC scattering with soft photons within the Nebula. The predicted fluxes of TeV $\gamma$-rays rely 
on the spatial distribution of the magnetic field within the Nebula as determined by the parameter $\sigma$ 
(ratio of energy density of magnetic field to the particle energy density) and/or by the average magnetic field 
$\rm <B>$ in the optical nebula. The HEGRA data are shown in Figure~6 together with predicted spectra using two 
SSC models of TeV $\gamma$-ray emission (de Jager et al. 1996; Atoyan \& Aharonian 1996). One may conclude that 
both models fit the HEGRA data rather well. 

According to the calculations of the TeV $\gamma$-ray emission by de Jager et al. (1996) the $\gamma$-ray flux 
from the Crab Nebula at TeV energies constrains the choice of the parameter $\sigma$. The IC spectrum computed 
by de Jager et al. (1996), assuming for the parameter $\sigma$ a value $\simeq 0.003$ gives a good fit to the 
HEGRA data. This value of the parameter $\sigma$ corresponds to the best-fitting magneto-hydrodynamic (MHD) 
solution of electron propagation in the Crab Nebula as found by Kennel \& Coroniti (1984). 

In another approach, assuming the spatial distribution of magnetic field in the Crab Nebula, one can determine 
the average magnetic field $\rm <B>$ in the optical nebula (Gould 1965). According to the calculations 
of Atoyan \& Aharonian (1996), made within the framework of the MHD model of Kennel \& Coroniti (1984), the 
average magnetic field $\rm <B>$ is determined by the TeV $\gamma$-ray flux as 
$\rm <B> \propto J_\gamma^{0.5}\cdot 10^{-5}\, G$ where $\rm J_\gamma,\, ph\, cm^{-2}s^{-1}TeV^{-1}$ is a 
differential flux at 1~TeV. Thus the HEGRA spectrum gives an average magnetic field strength  
$\rm <B> \simeq (1.7\pm 0.3) \cdot 10^{-4}G$. This value is consistent with the estimate derived by 
Hillas et al. (1998) from the Crab Nebula date taken with the 10~m Whipple telescope.

The IC energy spectrum of $\gamma$ rays from the Crab Nebula, measured in the energy range from 1~TeV to 10~TeV, 
is likely to be a power law $\rm dJ_\gamma /dE \propto E^{-\alpha}=E^{-2.6}$. At the same time, for the energy 
range above 10~TeV calculations predict gradual steepening with $\rm \alpha \simeq$2.7 and 2.9 at 10 and 30~TeV, 
respectively. 
That is due to both the energy loss of the ultra high energy electrons
by fast synchrotron cooling and the Klein-Nishina effect in the cross-section of the inverse Compton scattering.
Atoyan \& Aharonian (1996) and Bednarek \& Protheroe (1997) have shown that $\pi^0$-decay $\gamma$-ray fluxes, 
due to the relativistic protons accelerated in the Crab Nebula, may noticeably contribute at energies 
above 10~TeV. However the HEGRA Crab Nebula data expanded up to 20~TeV are still consistent with the pure IC 
spectrum. To assess the contribution of $\pi^0$-produced $\gamma$-rays from the Crab~Nebula measurements 
above 30~TeV are needed. Note that the LZA technique could help to perform such observations. 

The predicted IC $\gamma$-ray spectrum of the Crab Nebula is rather flat in the energy range below 
$\sim 1$~TeV. It could be well approximated by $\rm dJ_\gamma /dE \propto E^{-2.0}$ at 100~GeV. Detection 
of a gradual flattening in this energy range will prove the SSC scenario of TeV $\gamma$-ray emission. However 
the low energy points at the HEGRA Crab Nebula spectra ($\rm E_\gamma < 1\, TeV$) are strongly affected by 
possible systematic errors ($\simeq 50$\%) and do not allow such a conclusion. Future observations of the Crab 
Nebula with the forthcoming low threshold high sensitivity \v{C}erenkov detectors (see Catanese \& Weekes 1999) 
will offer precise measurements in these energy range.   
 
\section*{Acknowledgments}

The support of the German ministry for Research and technology BMBF and of the Spanish 
Research Council CYCIT is gratefully acknowledged. We thank the Instituto de Astrophysica 
de Canarias for the use of the site and for supplying excellent working conditions 
at La~Palma. We gratefully acknowledge the technical support staff of the Heidelberg, 
Kiel, Munich, and Yerevan Institutes.

\clearpage

\figcaption[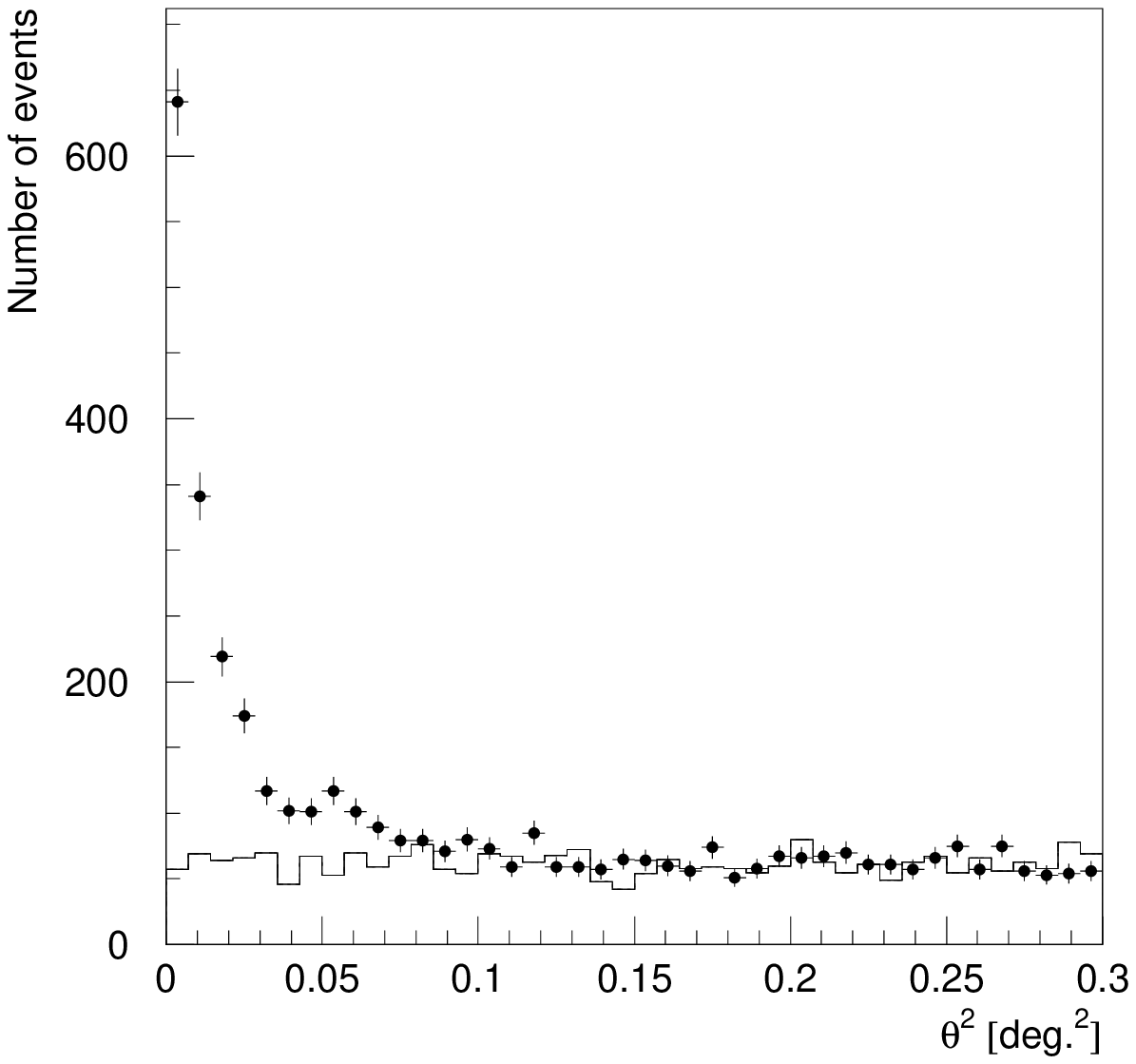]{Distribution over $\theta^2$ ($\theta$ is an angular 
distance of the true source position to the reconstructed source position) of ON (solid line) and OFF 
(dashed line) events in 12~hrs of observations of the Crab Nebula with the HEGRA IACT system at 
zenith angles $\leq 20^\circ$.}  

\figcaption[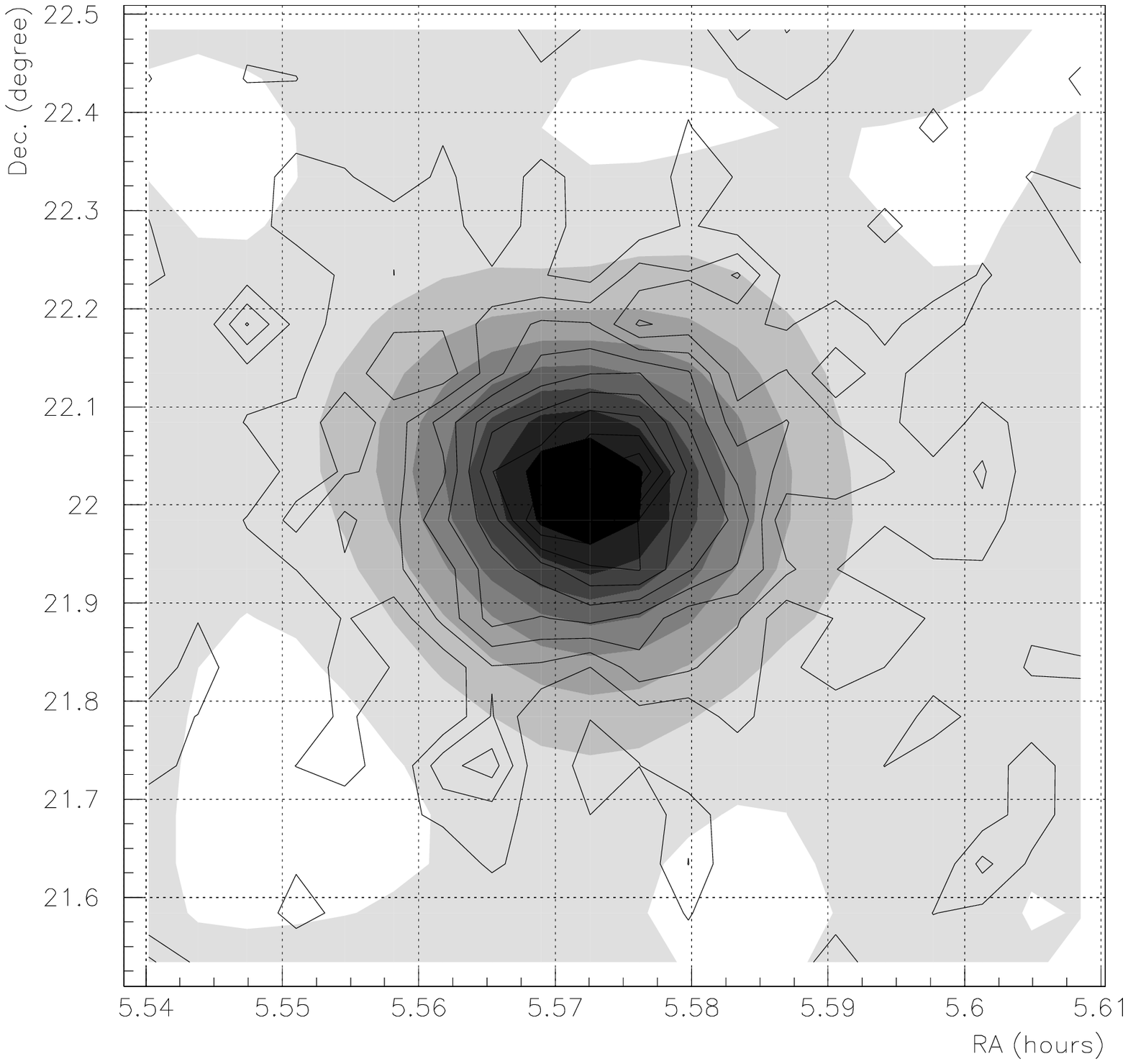]
{Filled contour plot (gray scale) of the smoothed distribution of the
reconstructed core shower positions in the sky (Dec. vs RA) in a 
$1^{o} \times 1^{o}$ region centered around the position of the Crab.
Superimposed are contours of the significance (the contours increment by 
one $\sigma$). Number of total events: 27800; number of events in the peak: 180. 
All events have $<\tilde{w}> < 1.2$.}

 
\figcaption[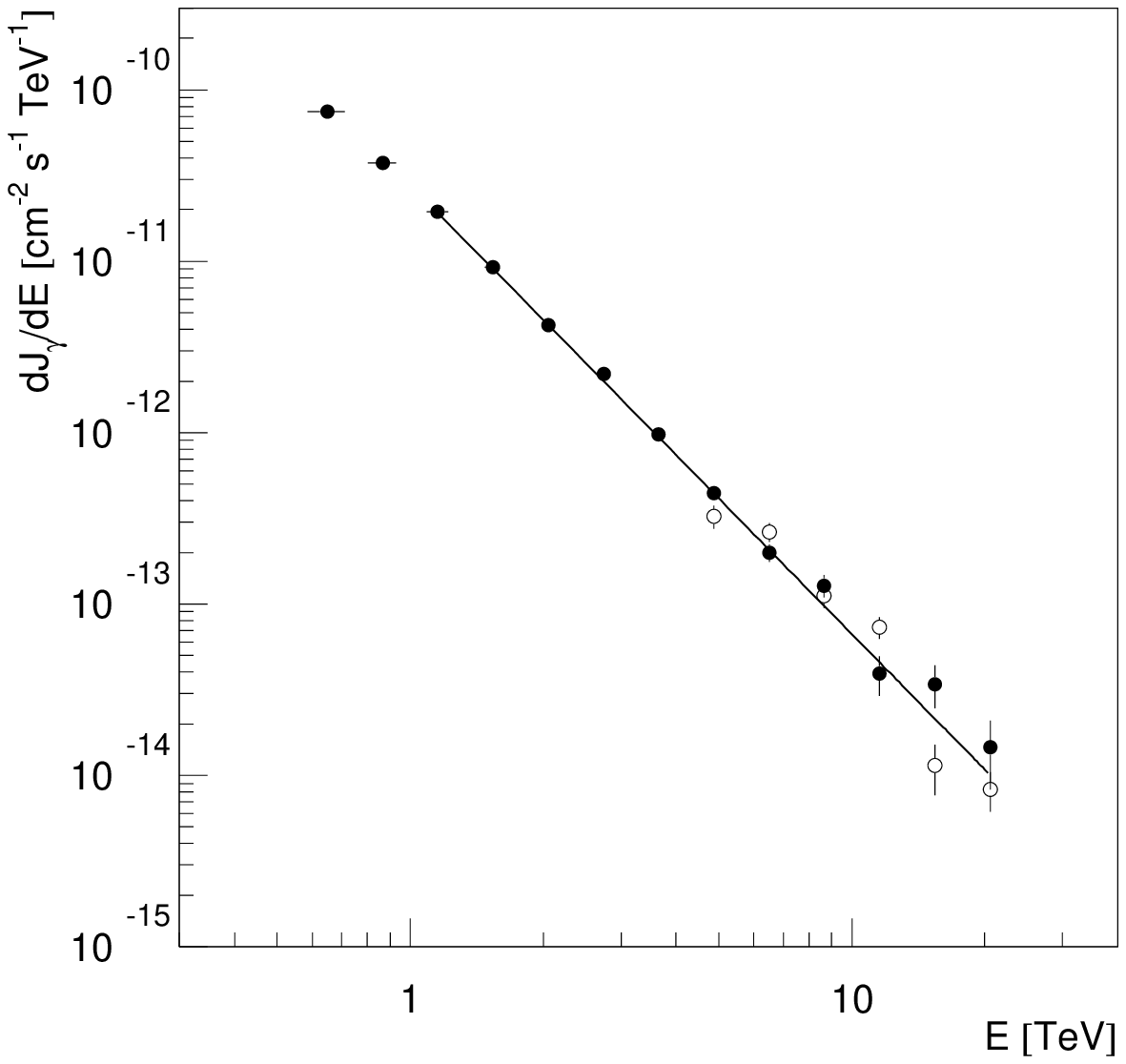]{The Crab~Nebula energy spectrum as measured at small and large 
zenith angles using the HEGRA IACT system. The filled circles are for the observations at zenith 
angles up to 50$^\circ$, the open circles are for the LZA data (60$^\circ$).
\label{fig1}}

\figcaption[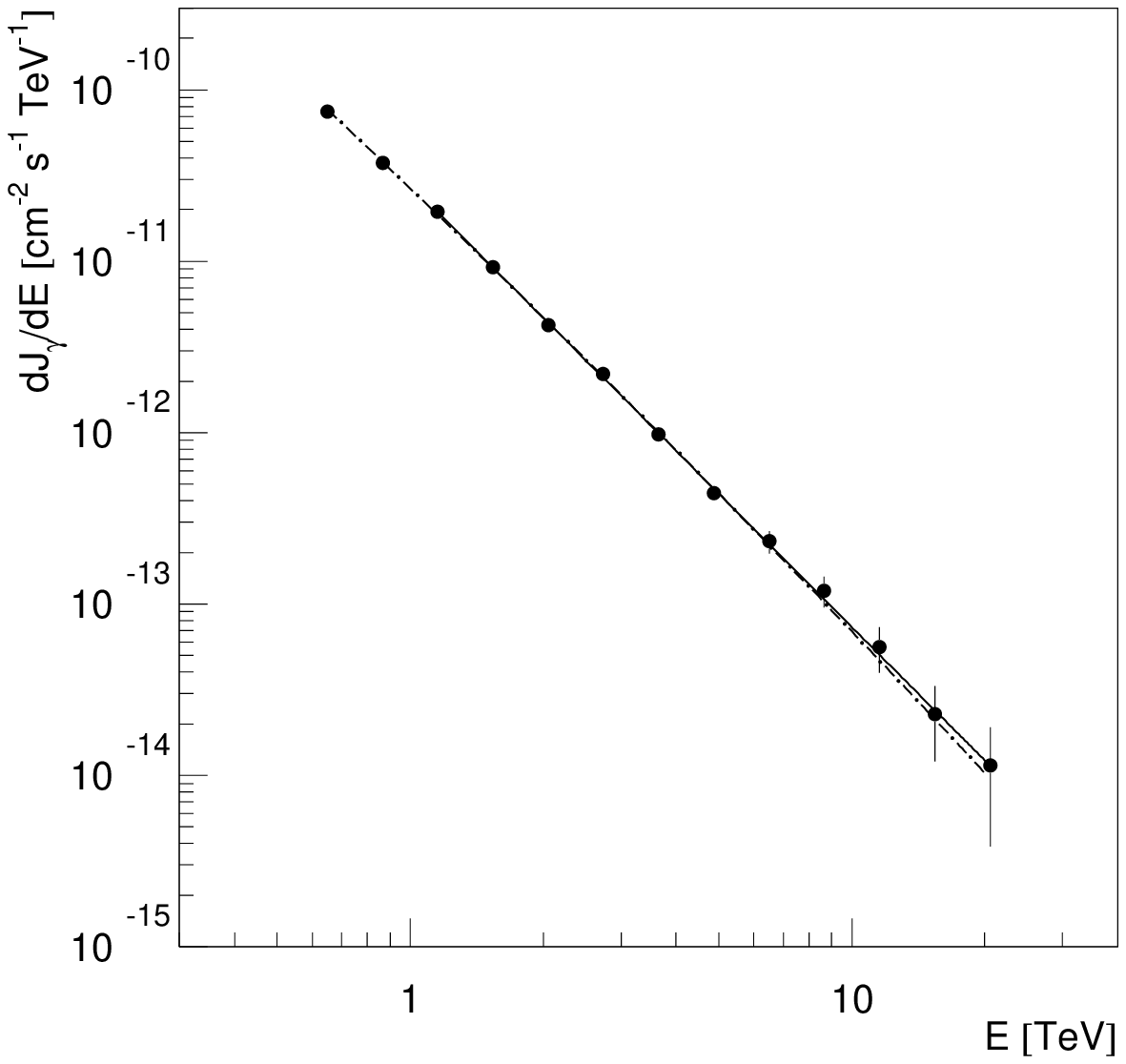]{The Crab~Nebula energy spectrum derived from the data taken over 
the zenith angle range from $0^\circ$ to $65^\circ$. The power law fit of eqn. (4) is shown 
by the solid line. The power law fit with logarithmically energy dependent slope of eqn. (5) is 
shown by the dashed curve.  
\label{fig2}}

\figcaption[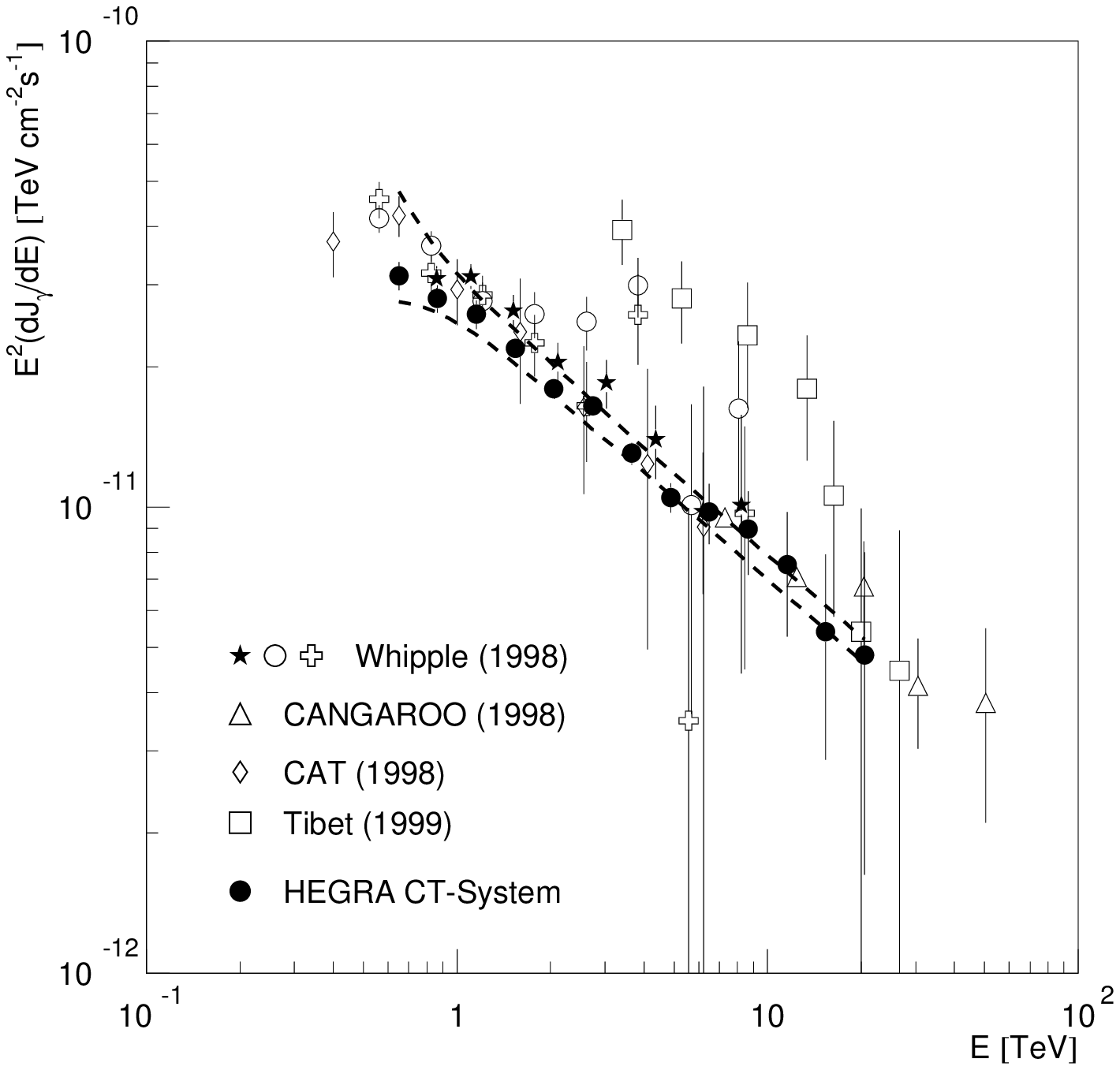]{Differential spectrum of the Crab Nebula as measured by the HEGRA 
system of imaging air \v{C}erenkov telescopes in comparison with the results of 
Whipple (Hillas~et~al.~1998), CANGAROO (Tanimori~et~al.~1998), CAT (Barrau 1998), and Tibet 
(Amenomori~et~al.~1999). The dashed curves show the upper and lower limits of the systematic 
error estimated for the HEGRA data.}

\figcaption[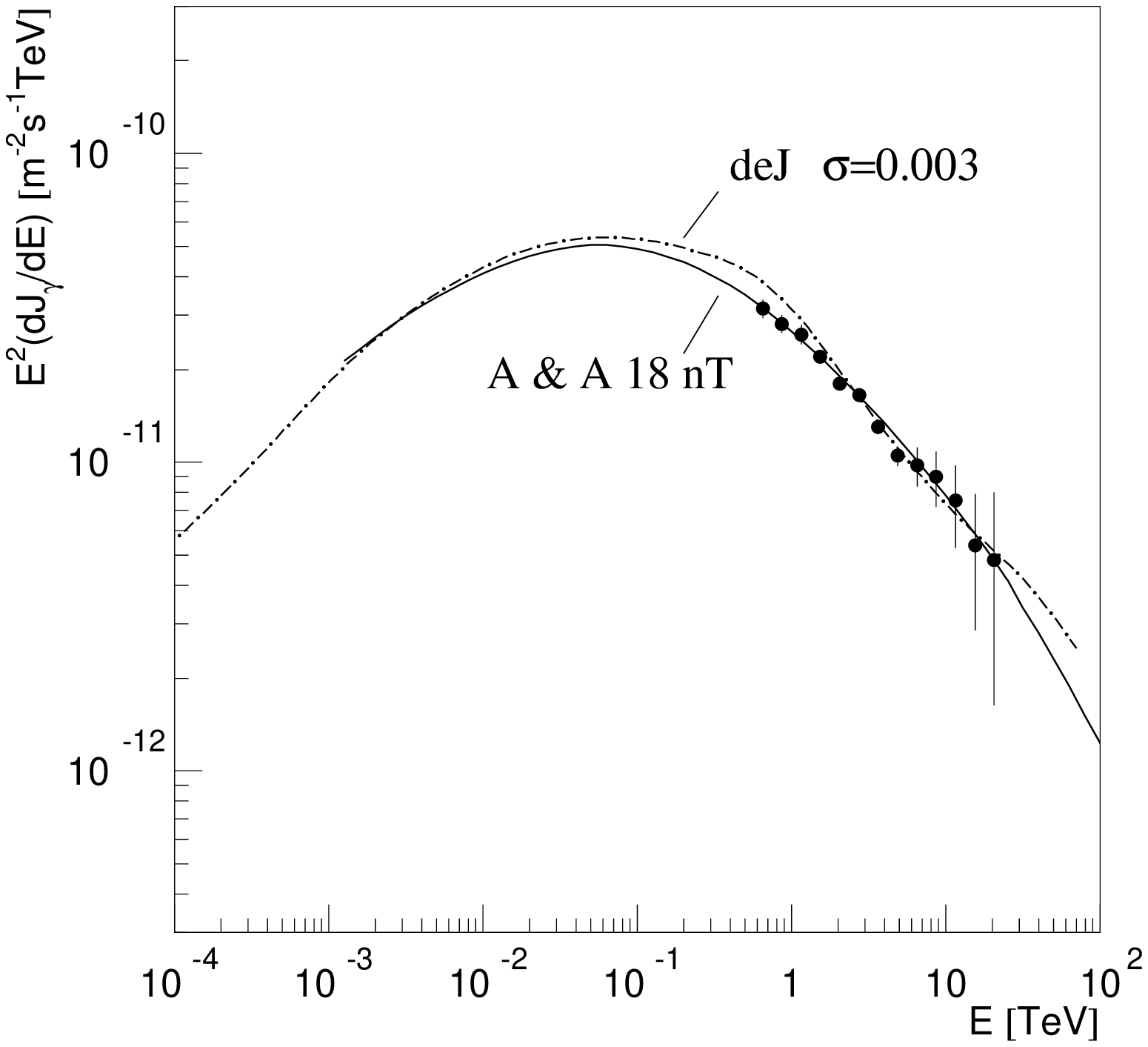]{
Spectral energy distribution of TeV $\gamma$-ray emission from  Crab~Nebula. The HEGRA data 
are shown by dots. Results of calculations using SSC model of emission 
made by Atoyan \& Aharonian (1996) and de Jager et al. (1996) are shown by solid and 
dotted-dashed curve, respectively.
\label{fig3}}

\clearpage

\begin{deluxetable}{clcll}
\footnotesize
\tablecaption{Summary of exposure times.}
\tablewidth{0pt}
\tablehead{
\colhead{N} & 
\colhead{Period}   & 
\colhead{Number of tel-s}   &
\colhead{Z.A.} &
\colhead{Time [hrs]} 
}
\startdata
1 & 1997 Sep - 1998 Jan        & 4    & $0^\circ - 50^\circ$     & 45.6 \\
2 & 1997 Oct - Nov           & 3    & $0^\circ - 50^\circ$     & 14.3 \\
3 & 1998 Feb - March         & 4    & $0^\circ - 50^\circ$     & 22.6 \\ \hline
  & Total:                   &      &                          & 82.5 \\ \hline
4 & 1998 Sep - Oct           & 4    & $0^\circ - 50^\circ$     & 26.2 \\  
  &                          &      & $50^\circ - 65^\circ$    & 5.4  \\ 
5 & 1998 Oct - Nov           & 5    & $50^\circ - 65^\circ$    & 29.8 \\
  &                          &      & $50^\circ - 65^\circ$    & 10.4 \\ 
6 & 1998 Dec                 & 5    & $0^\circ - 50^\circ$     & 20.1 \\ 
  &                          &      & $50^\circ - 65^\circ$    & 8.1  \\ \hline
  & Total:                   &      & $0^\circ - 50^\circ$     & 76.1 \\
  &                          &      & $50^\circ - 65^\circ$    & 24.0 \\ \hline 
\enddata
\end{deluxetable}


\begin{deluxetable}{clllll}
\footnotesize
\tablecaption{Summary of the data. No. of $\gamma$-rays, cosmic rays (CRs), and the S/N ratio,
in the standard deviations ($\sigma$).  
\label{tbl-2}}
\tablewidth{0pt}
\tablehead{
\colhead{ } & 
\colhead{Z.A.}   & 
\colhead{$0^\circ - 25^\circ$}  &
\colhead{$25^\circ - 40^\circ$} &
\colhead{$40^\circ - 50^\circ$} &
\colhead{$55^\circ - 65^\circ$} 
}
\startdata
         & Time, hrs & 7.30 & 11.34 & 7.56 & 5.44 \\ \hline
         & Raw       & 592 & 972 & 327 & 143 \\ 
No. of $\gamma$'s & L. cut    & 577 & 670 & 244 & 124 \\    
         & MD cut    & 450 & 499 & 157 & 101 \\ \hline
         & Raw       & 2964& 4495& 2772& 3353 \\
No. of CRs       & L. cut    & 536 & 846 & 567 & 658  \\ 
         & MD cut    & 191 & 268 & 151 & 114  \\ \hline
         & Raw       & 7.3 & 9.7 & 4.3 & 1.8  \\
S/N [$\sigma$] & L. cut & 14.2 & 13.8 & 6.6 & 3.3 \\
         & MD cut    & 15.6 & 15.5 & 7.3 & 5.6 \\ \hline 
\enddata
\tablecomments{
Data for the system of 4 IACTs were analyzed using a orientation cut of 
$\theta^2 \le 0.05 \rm \, deg^2$, and a loose mean scaled Width cut of 
$\rm <\tilde w> < 1.2$ (L. cut), a Mahalanobis distance cut (MD$\le 1.5$) 
         as well as a cut without image-shape (Raw).}
\end{deluxetable}

\begin{deluxetable}{clcll}
\footnotesize
\tablecaption{The integral $\gamma$-ray flux above 1~TeV, $\rm J_{\gamma}(>1\,TeV)$, in units of   
              $\rm photon \, cm^{-2} s^{-1}$, 
              measured during the different observation periods. $\rm R_{exp}$ and 
              $\rm R_{MC}$ are the measured and calculated hardware detection rates, respectively. 
              The data correspond to a zenith angle range from 0 to 50 degrees.
              \label{tbl-1}}
\tablewidth{0pt}
\tablehead{
\colhead{Period}   & 
\colhead{$\rm R_{exp}$}   &
\colhead{$\rm R_{MC}$} &
\colhead{$\rm J_{\gamma}$} 
}
\startdata
1 & 12.6 & 14.0 & $(1.68\pm 0.04)\cdot 10^{-11}$ \\
2 & 10.3 & 11.2 & $(1.71\pm 0.05)\cdot 10^{-11}$ \\
3 & 10.7 & 10.0 & $(1.68\pm 0.05)\cdot 10^{-11}$ \\ 
4 & 11.5 & 11.4 & $(1.77\pm 0.04)\cdot 10^{-11}$ \\  
5 & 10.1 & 10.0 & $(1.84\pm 0.04)\cdot 10^{-11}$ \\
6 & 11.8 & 11.4 & $(1.64\pm 0.05)\cdot 10^{-11}$ \\ 
\enddata
\tablecomments{These results were obtained using the power law fitting of the differential 
               $\gamma$-ray fluxes measured in the energy range from 1 to 10~TeV for 
               each observational period.}  
\end{deluxetable}

\clearpage

\begin{deluxetable}{clcll}
\footnotesize
\tablecaption{The energy spectrum of the Crab Nebula, $\rm dJ_\gamma/dE$, in unites of $\rm photon \, cm^{-2}s^{-1}$. 
              The statistical ($\rm \sigma_{stat}$) and systematic ($\rm \sigma_{stat}$) errors shown in 
              the table are also given in $\rm photon \, cm^{-2}s^{-1}$.}
\tablewidth{0pt}
\tablehead{
\colhead{E, TeV} &
\colhead{$\rm dJ_\gamma/dE$}  &
\colhead{$\sigma_{\rm stat}$} &
\colhead{$\sigma_{\rm syst}$} 
}
\startdata
0.65  & 7.44$\cdot 10^{-11}$ & 5.25$\cdot 10^{-12}$ & (+2.31-1.79)$10^{-11}$ \\
0.87  & 3.74$\cdot 10^{-11}$ & 2.50$\cdot 10^{-12}$ & (+5.98-5.61)$10^{-12}$ \\
1.16  & 1.95$\cdot 10^{-11}$ & 1.39$\cdot 10^{-12}$ & (+1.94-1.94)$10^{-12}$ \\
1.54  & 9.25$\cdot 10^{-12}$ & 1.58$\cdot 10^{-13}$ & (+8.32-7.40)$10^{-13}$ \\
2.05  & 4.25$\cdot 10^{-12}$ & 1.03$\cdot 10^{-13}$ & (+3.40-2.97)$10^{-13}$ \\
2.74  & 2.20$\cdot 10^{-12}$ & 7.72$\cdot 10^{-14}$ & $(\pm 1.54)10^{-13}$   \\
3.65  & 9.78$\cdot 10^{-13}$ & 5.17$\cdot 10^{-14}$ & $(\pm 5.87)10^{-14}$   \\
4.87  & 4.43$\cdot 10^{-13}$ & 3.25$\cdot 10^{-14}$ & $(\pm 2.66)10^{-14}$   \\
6.49  & 2.32$\cdot 10^{-13}$ & 3.40$\cdot 10^{-14}$ & $(\pm 1.39)10^{-14}$   \\
8.66  & 1.20$\cdot 10^{-13}$ & 2.45$\cdot 10^{-14}$ & $(\pm 7.19)10^{-15}$   \\
11.55 & 5.64$\cdot 10^{-14}$ & 1.68$\cdot 10^{-14}$ & $(\pm 3.38)10^{-15}$   \\  
15.40 & 2.28$\cdot 10^{-14}$ & 1.07$\cdot 10^{-14}$ & $(\pm 1.36)10^{-15}$   \\
20.54 & 1.14$\cdot 10^{-14}$ & 7.55$\cdot 10^{-15}$ & $(\pm 6.85)10^{-16}$   \\
\enddata

\end{deluxetable}

\clearpage

\begin{figure}[t]
\begin{center}
\includegraphics[width=0.8\linewidth]{crab2000.fig1.ps}
\end{center}
\end{figure}

\clearpage

\begin{figure}[t]
\begin{center}
\includegraphics[width=0.8\linewidth]{crab2000.fig2.ps}
\end{center}
\end{figure}

\clearpage

\begin{figure}[t]
\begin{center}
\includegraphics[width=1.0\linewidth]{crab2000.fig3.ps}
\end{center}
\end{figure}

\clearpage

\begin{figure}[t]
\begin{center}
\includegraphics[width=1.0\linewidth]{crab2000.fig4.ps}
\end{center}
\end{figure}

\clearpage

\begin{figure}[t]
\begin{center}
\includegraphics[width=0.9\linewidth]{crab2000.fig5.ps}
\end{center}
\end{figure}

\clearpage

\begin{figure}[t]
\begin{center}
\includegraphics[width=1.0\linewidth]{crab2000.fig6.ps}
\end{center}
\end{figure}

\end{document}